\documentclass[aps,amssymb,amsmath,prl,reprint,noshowpacs,superscriptaddress]{revtex4-2}
\usepackage[utf8]{inputenc}
\usepackage{amsthm}
\usepackage{mathtools}
\usepackage{physics}
\usepackage{xcolor}
\usepackage{graphicx}
\usepackage[left=23mm,right=13mm,top=35mm,columnsep=15pt]{geometry} 
\usepackage{adjustbox}
\usepackage{placeins}
\usepackage[T1]{fontenc}
\usepackage{lipsum}
\usepackage{csquotes}
\usepackage{siunitx}
\usepackage[english]{babel}
\usepackage{soul}
\usepackage{color}
\usepackage{helvet}
\usepackage{scalefnt} 
\usepackage{natbib}
\usepackage{hyperref}
\usepackage{appendix}
\usepackage{color}
\usepackage{comment}
\hypersetup{
	colorlinks   = true, 
	urlcolor     = blue, 
	linkcolor    = blue,
	citecolor   = blue 
}
\begin{document}

\scalefont{1.0}

\title{Cavity-mediated long-range interactions in levitated optomechanics}

\author{Jayadev~\surname{Vijayan}}
\thanks{Equal contribution}
\email{jvijayan@ethz.ch}
\affiliation{Photonics Laboratory, ETH Z{\"u}rich, Z\"urich, Switzerland}
\affiliation{Quantum Center, ETH Z{\"u}rich, Z\"urich, Switzerland}
\author{Johannes~\surname{Piotrowski}}
\thanks{Equal contribution}
\affiliation{Photonics Laboratory, ETH Z{\"u}rich, Z\"urich, Switzerland}
\affiliation{Quantum Center, ETH Z{\"u}rich, Z\"urich, Switzerland}
\author{Carlos~\surname{Gonzalez-Ballestero}}
\affiliation{Institute for Theoretical Physics, University of Innsbruck, Innsbruck, Austria}
\affiliation{Institute for Quantum Optics and Quantum Information, Austrian Academy of Sciences, Innsbruck, Austria}
\author{Kevin~\surname{Weber}}
\affiliation{Photonics Laboratory, ETH Z{\"u}rich, Z\"urich, Switzerland}
\affiliation{Quantum Center, ETH Z{\"u}rich, Z\"urich, Switzerland}
\author{Oriol~\surname{Romero-Isart}}
\affiliation{Institute for Theoretical Physics, University of Innsbruck, Innsbruck, Austria}
\affiliation{Institute for Quantum Optics and Quantum Information, Austrian Academy of Sciences, Innsbruck, Austria}
\author{Lukas~\surname{Novotny}}
\affiliation{Photonics Laboratory, ETH Z{\"u}rich, Z\"urich, Switzerland}
\affiliation{Quantum Center, ETH Z{\"u}rich, Z\"urich, Switzerland}


\begin{abstract}

The ability to engineer cavity-mediated interactions has emerged as a powerful tool for the generation of non-local correlations and the investigation of non-equilibrium phenomena in many-body systems.
Levitated optomechanical systems have recently entered the multi-particle regime, with promise for using arrays of massive strongly coupled oscillators for exploring complex interacting systems and sensing.
Here, by combining advances in multi-particle optical levitation and cavity-based quantum control, we demonstrate, for the first time, programmable cavity-mediated interactions between nanoparticles in vacuum.
The interaction is mediated by photons scattered by spatially separated particles in a cavity, resulting in strong coupling ({$G_\text{zz}/\Omega_\text{z} = 0.238\pm0.005$}) that does not decay with distance within the cavity mode volume.
We investigate the scaling of the interaction strength with cavity detuning and inter-particle separation, and demonstrate the tunability of interactions between different mechanical modes.
Our work paves the way towards exploring many-body effects in nanoparticle arrays with programmable cavity-mediated interactions, generating entanglement of motion, and using interacting particle arrays for optomechanical sensing. 

\end{abstract}

\maketitle

\label{sec:introduction}
\centerline{\textbf{Introduction}}
\vspace{0.7mm}

Exploring quantum physics at macroscopic scales is an exciting prospect, both for fundamental physics as well as developing technology~\cite{Leggett2002,Gonzalez-Ballestero2021, Rademacher2020}.
However, in addition to the challenge of ground-state cooling massive objects, such endeavours require either large-scale delocalization of a single object, or the entanglement of multiple objects.
Levitodynamics, which deals with controlling the mechanical motion of massive oscillators in vacuum~\cite{Millen2020,Gonzalez-Ballestero2021}, has made remarkable headway towards multiple-particle systems, with demonstrations of cooling~\cite{Vijayan2023} and short-range coupling~\cite{Rieser2022,Arita2022,Penny2023,Bykov2023,Brzobohaty2023,Liska2023} between nanoparticles levitated in free space.
Furthermore, recent experiments with single particles in optical tweezers have established exquisite control over rotational dynamics~\cite{Bang2020,vanderLaan2021,Pontin2023,Zielinska2023}, and achieved quantum ground-state cooling of mechanical motion~\cite{Delic2020,Magrini2021,Tebbenjohanns2021,Kamba2022,Ranfagni2022,Piotrowski2023}.
One of the next pivotal milestones towards macroscopic quantum physics is to 
entangle multiple particles via optical forces.
However, this is not possible in free space, as the entangling rate is not large enough to overcome the decoherence rates of the particles~\cite{Rudolph2020, Rudolph2023}.
Therefore, it becomes desirable to use an optical cavity to mediate coupling between the particles.

Here, we introduce such a capability to engineer strong programmable cavity-mediated interactions between multiple spatially separated particles in vacuum.
The programmability arises from the use of acousto-optic deflectors (AODs) to generate tweezer arrays for trapping the particles~\cite{Vijayan2023,Yan2023}, which offer a high degree of control over parameters such as optical frequencies of the tweezers, cavity detuning, as well as mechanical frequencies and position of the particles.
Such parameter control is crucial for precisely tuning the interaction strength and for choosing which particles and mechanical modes couple.

Most experimental systems that study many-body physics rely either on localized short-range interactions~\cite{Jurcevic2014,Bernien2017}, or a common cavity mode to mediate interactions~\cite{Leroux2010,Pedrozo2020}, and can only afford short-distance or all-to-all connectivity respectively.
In our platform, on the other hand, the decoupling of the trapping mechanism from the cavity presents the opportunity to engineer a much broader class of connectivity, by programming specific tweezers to be resonant with the cavity mode.
The prospect of using programmable cavity-mediated interactions in levitodynamics will facilitate progress towards synthesising arrays with tunable effective geometry and connectivity~\cite{Kollar2019,Periwal2021}, generating quantum correlations and entanglement~\cite{Chauhan2020,Brandao2021,Kotler2021,Lepinay2021,Rudolph2020,Weiss2021, Chauhan2022}, exploring complex phases emerging from interacting particles~\cite{Reimann2015,Landig2016,Bernien2017,Liu2020} and sensing with mechanical sensor arrays~\cite{Monteiro2020,Carney2020,Brady2022}.

\vspace{4mm}

\label{sec:setup}
\centerline{\textbf{Experimental setup}}
\vspace{0.7mm}

\begin{figure*}
    \centering
    \includegraphics[width = 18cm]{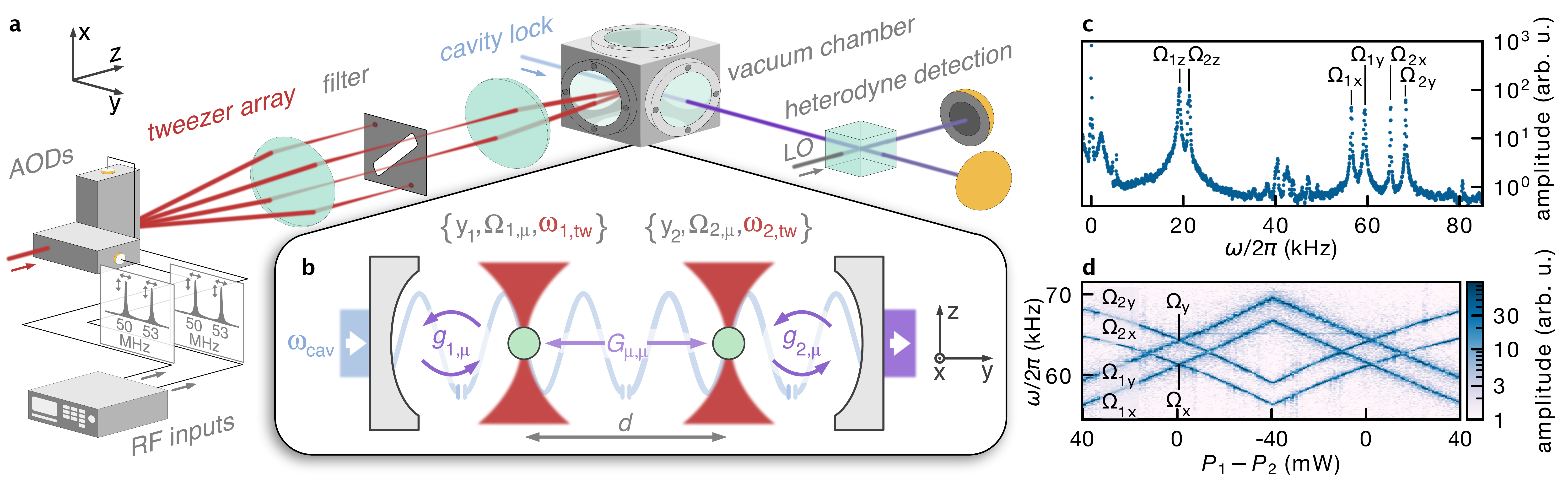}
    \caption{\textbf{Cavity optomechanics with multiple nanoparticles.} 
    \textbf{a},
    ~ A laser beam is split using acousto-optical deflectors (AODs) and focused by a high-NA focusing lens inside a vacuum chamber to generate optical tweezers.
    The optical frequency of the tweezers $\omega_\text{tw}$, as well as the positions $y_i$ of the particles ($i\in (1,2)$) and the mechanical frequencies $\Omega_{i,\mu}$ of their center-of-mass modes ($\mu\in (x,y,z)$), are controlled by programming the radiofrequency (RF) inputs of the AODs.
    \textbf{b},~The particles are positioned with a variable spatial separation $d =  |y_\text{1}-y_\text{2}|$ along the standing wave of an optical cavity, which is blue-detuned by $\Delta = \omega_\text{cav} - \omega_{\text{tw}}$.
    The individual optomechanical coupling of each particle with strength $g_{i,\mu}$ introduces an effective coupling $G_{\mu,\mu}$ between the two particles. 
    Tweezer light scattered by the particles leaks out of the cavity and is interfered with a local oscillator for balanced heterodyne detection.
    \textbf{c},~Spectrum of the heterodyne signal, where we set the Rayleigh peak frequency to $\omega=0$, showing the three mechanical anti-Stokes sidebands at $\omega = \Omega_{i,\mu}$ for each particle.
    \textbf{d},~Spectrogram of the anti-Stokes sidebands as a function of the power difference between the tweezers $P_1-P_2$.
    }
    \label{fig:setup}
\end{figure*}

The mechanical oscillators in our experiment comprise of near-spherical SiO$_2$ nanoparticles with nominal diameter $150\,$nm, levitated in vacuum ($\sim 10^{-4}\,$mbar) using optical tweezers (NA $ = 0.75$) at wavelength $\lambda = 1550\,$nm (see Fig.~\ref{fig:setup}a). 
The particles are placed in the TEM$_\text{00}$ mode of an optical cavity with linewidth $\kappa/2\pi = 600\,$kHz, comprised of mirrors with different transmissions separated by $9.6\,$mm (see Fig.~\ref{fig:setup}b).
We use two tweezers with identical optical frequencies along the diagonal of a two-dimensional array of beams generated by two orthogonally placed acousto-optical deflectors (AODs)~\cite{Endres2016}. 
The cavity resonance is detuned by a frequency $\Delta$ with respect to both tweezers. 
Light scattered by the particles, carrying information about their center-of-mass motion along the three axes $x$, $y$ and $z$, leaks through the higher transmission cavity mirror.
This light is combined with a local oscillator and split equally onto a balanced photodetector. 
The Fourier transform of the detector voltage gives the spectral amplitudes of our signal.
For convenience, we offset the frequency axis to have the optical tweezer frequency at zero (see Fig.~\ref{fig:setup}c).
\begin{figure*}
    \centering
    \includegraphics[width = 18cm]{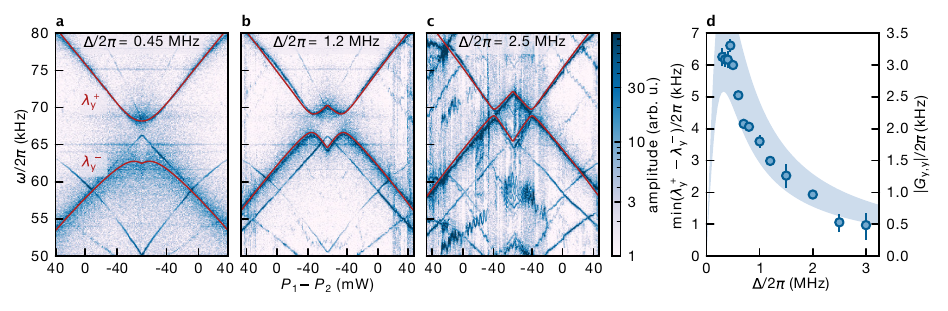}
    \caption{\textbf{Cavity-mediated long-range interactions.}
    \textbf{a-c},~Measured spectrograms show the normal mode splittings arising from cavity-mediated particle-particle interactions for different values of cavity detuning $\Delta$. 
    Red lines are fits of the normal mode frequencies $\lambda_y^-$ and $\lambda_y^+$ of the coupled system. 
    \textbf{d},~The splitting $\min(\lambda_y^+-\lambda_y^-) = 2 |G_{y,y}|$ of the avoided crossing extracted from the fits, as a function of detuning. Error bars correspond to three s.d. of the fitted coupling strengths. 
    The shaded area shows theoretical estimations of the coupling strength $G_{y,y}$ based on system parameters, exhibiting the characteristic dependence on cavity detuning.
    }
    \label{fig:LRI}
\end{figure*}
The particle positions $y_i$ ($i\in (1,2)$) and their separation $d = |y_\text{1}-y_\text{2}|$ along the cavity axis are controlled by the frequencies sent to the radio-frequency (RF) channels of the AODs, while preserving the degeneracy of the tweezer optical frequencies (see Supplementary). 
In addition, a translation stage for the high-NA lens can displace both particles simultaneously. 
The spatial separation $d$ between the nanoparticles ({\color{black}typically $6\,\mathrm{\mu}$m}) is large enough to suppress short-range Coloumb ($\propto 1/d^3$)~\cite{Rudolph2022,Penny2023} and free-space optical binding ($\propto 1/d$)~\cite{Rieser2022} interactions.
The optical power of the tweezers $P_i$ (typically $130\,$mW) and resulting mechanical frequencies $\Omega_{i,\mu}$ ($\mu\in (x,y,z)$) are set by the respective AOD RF amplitudes. 

To engineer interactions in our experiments, we bring the mechanical frequencies of the two particles close together by carrying out linear ramps of the optical powers. 
We make use of coherent scattering, whereby light scattered off each nanoparticle populates the cavity mode and results in optomechanical coupling with strengths $g_{i,\mu}$~\cite{GonzalezBallestero2019}. 
This gives rise to effective cavity-mediated interactions between the nanoparticles. 
For particles in tweezers polarised along the cavity ($y$) axis, $g_{i,\mu}$ is zero and there is no particle-particle coupling ($G_{\mu,\mu}\propto g_{\text{1},\mu}g_{\text{2},\mu} = 0$).
Therefore, a power ramp simply results in crossings of the mechanical frequencies at $\Omega_\mu$, as seen in the spectrogram in Fig.~\ref{fig:setup}d.
In the following, the tweezers are polarized along the $x$-axis, maximising light scattering along the cavity axis. 
The couplings along the $x-$axis are minimal in this configuration, but the couplings along $y-$ and $z-$axis can be maximal and result in cavity-mediated long-range interactions.

\vspace{4mm}

\label{sec:cooling}
\centerline{\textbf{Cavity-mediated long-range interactions}}
\vspace{0.7mm}

Two mechanical modes of different nanoparticles coupled to the same cavity mode via coherent scattering gives rise to an effective particle-particle coupling
\begin{equation}
G_{\mu,\mu}=\frac{g_{\text{1},\mu} (g_{\text{2},\mu})^*}{\left(\Delta+\Omega_\mu\right)+i \kappa / 2}+\frac{(g_{\text{1},\mu})^* g_{\text{2},\mu}}{\left(\Delta-\Omega_\mu\right)-i \kappa / 2}\;,
    \label{eq:g_eff}
\end{equation}
when the mechanical frequencies $\Omega_{\text{1},\mu} \approx \Omega_{\text{2},\mu}$ are close to degeneracy ($\approx \Omega_\mu$). 
Details are provided in the supplementary.
The coupled system has two normal modes with frequencies $\lambda_\mu^-$ and $\lambda_\mu^+$. 
The minimal splitting at the avoided crossing between these normal modes is $\min(\lambda_\mu^+-\lambda_\mu^-) = 2 |G_{\mu,\mu}|$.
In Fig.~\ref{fig:LRI}a,b,c we show spectrograms of the $x$ and $y$ modes of two particles positioned at separate cavity nodes during power sweeps for $\Delta/2\pi = 0.45, 1.2 , 2.5 \,$MHz, respectively.
As in Fig.~\ref{fig:setup}d, the $x$ modes cross at $\Omega_x/2\pi \sim 59\,$kHz, indicating no interactions of the $x$ degree of freedom and offering a calibration of optical powers through $\Omega_{i,x} \propto \sqrt{P_i}$.
We fit expressions for $\lambda_y^-$ and $\lambda_y^+$ (see Supplementary) to the normal modes in the spectrograms using the calibrated powers and the bare mechanical frequencies $\Omega_{i,y}$ at the edges of the ramp, with the only free parameter being the product of the individual optomechanical coupling strengths $|g_{\text{1},y} g_{\text{2},y}|$.
We overlay the fits in Fig.~\ref{fig:LRI}a-c and extract $\min(\lambda_y^+-\lambda_y^-)$, which follows the characteristic dependence on the detuning $\Delta$, given by Eq.~\ref{eq:g_eff}.

We see two other features of cavity optomechanics as the cavity is brought closer to resonance with the mechanical frequencies of the particles.
First, the $y$ modes of both particles are cooled via coherent scattering~\cite{Windey2019,Piotrowski2023} and become visibly broader in the spectrograms, whereas the $x$ modes are not.
Second, the `cavity-pulling' effect~\cite{Sommer2021} causes a larger shift of the mechanical frequencies of the $y$ modes as the detuning decreases. 
Therefore, the avoided crossings of the $y$ modes appear shifted away from the crossings of the $x$ modes at $P_1=P_2$.
Fig.~\ref{fig:LRI}d shows the normal mode splitting over a wide range of detunings.
The shaded area represents the particle-particle coupling strength calculated from Eq.~\ref{eq:g_eff}, with $g_{\text{1},\mu}$ and $g_{\text{2},\mu}$ estimated from system parameters and their uncertainties (see Supplementary). The measured splittings are in excellent agreement with the theoretical estimates.
We observe a maximum splitting of $\min(\lambda_y^+-\lambda_y^-)/2\pi = (6.6\pm0.2)\,$kHz at $\Delta/2\pi = 0.45\,$MHz, close to the optimal detuning.

\vspace{4mm}

\label{sec:noneq}
\centerline{\textbf{Distance-dependence of interactions}}
\vspace{0.7mm}

\begin{figure}
    \includegraphics[width = 9cm]{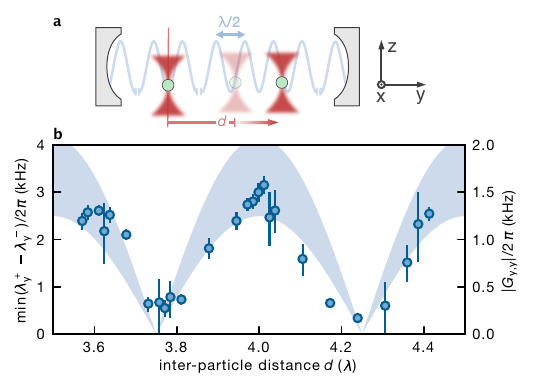}
    \caption{\textbf{Distance-dependence of cavity-mediated interactions.}
    \textbf{a},~Sketch of two particles placed in the standing wave of the cavity. 
    The position of the first particle (left) is kept fixed at a node whereas the second particle (right) is moved along the standing wave, thereby increasing the inter-particle distance $d$.
    \textbf{b},~Data points show the measured mode splitting as a function of $d$ with error bars corresponding to three s.d. of the fitted coupling strengths. 
    The shaded area shows the position dependence of $g_{\text{2},y}\propto |\cos{(2\pi d/\lambda)}|$ imprinted on the particle-particle coupling $G_{y,y}$. 
    Its width corresponds to uncertainties in system parameters. 
    }
    \label{fig:collapseandrevival}
\end{figure}

\begin{figure*}
    \centering
    \includegraphics[width = 18cm]{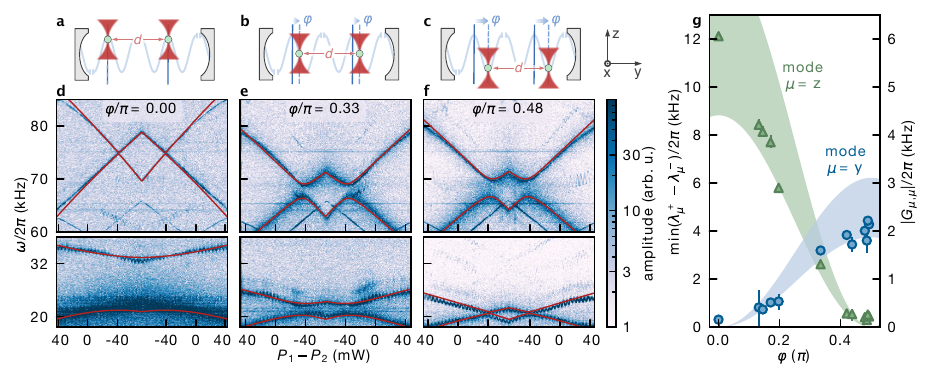}
    \caption{\textbf{Tunability of interactions between mechanical modes.}
    By keeping the inter-particle separation $d$ fixed and moving the particle pair along the cavity axis, we can tune the interaction strengths of different mechanical modes. 
    The distance of each particle to the closest cavity node is characterised by the phase factor $\varphi$.
    \textbf{a,d},~Both particles are located at anti-nodes - the $z$ modes interact giving rise to significant mode splitting, whereas the $y$ modes show no sign of interactions (no mode splitting).
    \textbf{b,e},~Both particles are positioned along slopes of the standing wave - the $z$ mode splitting decreases and the $y$ modes start splitting. 
    \textbf{c,f},~Both particles are located at nodes - the $y$ modes show avoided crossing, whereas the splitting of the $z$ modes fades away.
    In all spectrograms, we fit the normal mode frequencies $\lambda_y^-$ and $\lambda_y^+$ (red lines).
    \textbf{g},~The splittings $\min(\lambda_z^+-\lambda_z^-)$ (green triangles) and $\min(\lambda_y^+-\lambda_y^-)$ (blue circles) as extracted from the fits with error bars of three s.d. of the fitted coupling strengths. 
    Strong $z$ coupling at the antinodes ($\varphi=0$) transitions to strong $y$ coupling at the nodes ($\varphi=\pi/2$). 
    Shaded areas show $G_{z,z}$ (green) and $G_{y,y}$ (blue) from system parameters and their uncertainties.
    }
    \label{fig:commonmodescan}
\end{figure*}

In addition to cavity-mediated interactions, the two particles can interact via Coloumb forces (provided they are charged) or via direct optical dipole-dipole coupling.
Although such interactions offer a useful resource in levitated optomechanics - e.g., for sympathetic cooling~\cite{Arita2022,Bykov2023, Penny2023}, engineered coupling~\cite{Rieser2022} and synchronization~\cite{Brzobohaty2023} - they are of short-range nature ($\propto1/d^3$ and $\propto1/d$ respectively). 
Furthermore, charged particles introduce electronic noise and the scattering losses of direct optical interactions prevent the generation of entanglement~\cite{Rudolph2023}.

By examining the dependence on inter-particle distance $d$, we show that the particle-particle interactions in our system are mediated via the cavity mode.
We keep one particle stationary at a node and scan the position of the second particle along the nodes and anti-nodes of the cavity standing wave (see Fig.~\ref{fig:collapseandrevival}a), while maintaining a fixed detuning of $\Delta/2\pi = 1.2\,$MHz. 
By briefly separating the optical frequencies of both tweezers and measuring the magnitude of the Rayleigh peaks in the heterodyne spectrum, we can independently determine the positions $y_\text{1}$ and $y_\text{2}$ of the two particles along the cavity standing wave (see Supplementary)~\cite{Windey2019,Piotrowski2023}. 
We perform optical power sweeps, as in the previous section, for different distances and extract normal mode splittings from the resulting spectrograms. 
Fig.~\ref{fig:collapseandrevival}b shows that as the position of the second particle is scanned along the cavity axis, the splitting exhibits a periodic behavior.
While the optomechanical coupling strength $g_{\text{1},y}$ is constant throughout this measurement (since $y_\text{1}$ is fixed), $g_{\text{2},y}$ follows a periodic dependence on $d$ (as we change $y_\text{2}$) and imprints it on the particle-particle coupling through $|G_{y,y}| \propto |g_{\text{2},y}| \propto |\cos{(2\pi d/\lambda)}|$.
The shaded area in Fig.~\ref{fig:collapseandrevival}b shows $G_{y,y}$ as calculated from system parameters using Eq.~\ref{eq:g_eff}, in agreement with the measured splittings. 
Conservative estimates of coupling strengths due to Coulomb ({\color{black}50 elementary charges} on each nanoparticle) and short-range optical interactions from our system parameters at {\color{black}$d/\lambda = 3.5$ give maximal coupling values of $G_\text{C}/2\pi = 0.17\,$kHz and $G_\text{O}/2\pi = 0.14\,$kHz, respectively (see Supplementary). 
Both are close to the resolution limit of our fitting procedure of $G_\mathrm{min}/2\pi \sim 0.15\,$kHz, given by the peak widths of $0.6\,$kHz}. 

\vspace{4mm}


\label{sec:simultaneous}
\centerline{\textbf{Tunability of interacting modes}}
\vspace{0.7mm}

The optomechanical coupling between each mode of the particle and the cavity is position dependent.
The positioning control in our setup allows us to vary the relative interaction strengths of the transverse $y$ mode and the longitudinal $z$ mode.
Previous studies have shown that the transverse coupling strengths scale as $g_{i,x},g_{i,y} \propto \sin{\varphi_i}$, and the longitudinal coupling strengths as $g_{i,z} \propto \cos{\varphi_i}$, where the phase factors $\varphi_i$ encode the distance between the particle position $y_i$ and the closest intensity maximum of the cavity~\cite{Windey2019,GonzalezBallestero2019}.

In our final experiment, we keep the inter-particle separation fixed at {\color{black}$d=4\lambda$}, such that $\varphi_1=\varphi_2=\varphi$, and move both particles simultaneously along the standing wave from anti-nodes ($\varphi = 0$) to nodes ($\varphi = \pi/2$) (see Fig.~\ref{fig:commonmodescan}a-c).
Doing so, we observe the cavity-mediated interactions transition from the $z$ modes  to the $y$ modes (Fig.~\ref{fig:commonmodescan}d-f).
Notably, at anti-nodes (Fig.~\ref{fig:commonmodescan}d), we observe a large normal mode splitting for the $z$ mode, corresponding to particle-particle coupling as high as $G_\text{zz}/\Omega_\text{z} = 0.238\pm0.005$.
The mode splittings as a function of phase between $\varphi = 0$ and $\varphi = \pi/2$ are shown in Fig.~\ref{fig:commonmodescan}g.
The shaded areas are the particle-particle coupling strengths estimated from system parameters and their uncertainties, exhibiting the dependencies $G_{y,y}\propto\sin^2{\varphi}$ and $G_{z,z}\propto\cos^2{\varphi}$. 
Therefore, the choice of the position of our particles allows us to precisely tune the relative interaction strengths of the different modes of the mechanical oscillator.

\vspace{4mm}

\label{sec:conclusions}
\centerline{\textbf{Conclusion and outlook}}
\vspace{0.5mm}

Combining the capabilities of cavity-based coherent scattering with multi-particle levitation provides a novel platform for optomechanics.
The high degree of control over parameters such as cavity detuning, mechanical frequency, polarization, and particle position has allowed us to engineer and investigate the nature of cavity-mediated long-range interactions between two mechanical oscillators. 
We investigated the scaling of the interactions strength of the transverse $y$ modes of two particles with cavity detuning, explored the distance-dependence of the interactions, and finally showed that we can tune the interactions of different mechanical modes of the two particles.  
The highest interaction strength we report is $G_\text{zz}/\Omega_\text{z} = 0.238\pm0.005$ for the longitudinal $z$ modes.
This value is higher than reported in free-space experiments~\cite{Rieser2022}, despite the much larger separations in our experiments.
Our scheme can be readily scaled up to more particles by splitting a higher-power laser into different paths to generate arrays of tweezers.
Additionally, switching to a cavity with smaller mode volume or a narrower linewidth~\cite{Sommer2021,Dare2023} can increase the interaction strength further to meet the requirements for motional entanglement~\cite{Rudolph2020}.

Ultimately, the ability to engineer programmable cavity-mediated interactions between levitated systems offers a powerful new resource in optomechanics.
Together with advances in achieving quantum control of mechanical motion and scaling up to nanoparticle arrays, our work firmly establishes levitodynamics as a compelling platform to explore the boundaries of quantum physics with massive interacting mechanical systems, and to build ultra-precise sensors with optomechanical arrays. 

\smallskip

\emph{Data availability:} The datasets generated and analysed during the current study will be made available in the ETH Zurich Research Collection prior to publication.

\smallskip 

\emph{Acknowledgements:} We thank our colleagues at the ETH Photonics Lab, L. Festa and A. Omran for insightful discussions.
This research was supported by the Swiss National Science Foundation (grant no. 215917, UeM019-2), European Research Council (grant no. 951234, Q-Xtreme), and European Union's Horizon 2020 research and innovation programme (grant no. 863132, iQLev).

\bibliographystyle{apsrev4-1}

\bibliography{1main}

\end{document}


\scalefont{1.0}

\title{Supplementary materials \\ Cavity-mediated long-range interactions in levitated optomechanics}

\author{Jayadev~\surname{Vijayan}}
\thanks{Equal contribution}
\email{jvijayan@ethz.ch}
\affiliation{Photonics Laboratory, ETH Z{\"u}rich, Z\"urich, Switzerland}
\affiliation{Quantum Center, ETH Z{\"u}rich, Z\"urich, Switzerland}
\author{Johannes~\surname{Piotrowski}}
\thanks{Equal contribution}
\affiliation{Photonics Laboratory, ETH Z{\"u}rich, Z\"urich, Switzerland}
\affiliation{Quantum Center, ETH Z{\"u}rich, Z\"urich, Switzerland}
\author{Carlos~\surname{Gonzalez-Ballestero}}
\affiliation{Institute for Theoretical Physics, University of Innsbruck, Innsbruck, Austria}
\affiliation{Institute for Quantum Optics and Quantum Information, Austrian Academy of Sciences, Innsbruck, Austria}
\author{Kevin~\surname{Weber}}
\affiliation{Photonics Laboratory, ETH Z{\"u}rich, Z\"urich, Switzerland}
\affiliation{Quantum Center, ETH Z{\"u}rich, Z\"urich, Switzerland}
\author{Oriol~\surname{Romero-Isart}}
\affiliation{Institute for Theoretical Physics, University of Innsbruck, Innsbruck, Austria}
\affiliation{Institute for Quantum Optics and Quantum Information, Austrian Academy of Sciences, Innsbruck, Austria}
\author{Lukas~\surname{Novotny}}
\affiliation{Photonics Laboratory, ETH Z{\"u}rich, Z\"urich, Switzerland}
\affiliation{Quantum Center, ETH Z{\"u}rich, Z\"urich, Switzerland}

\date{\today} 
\maketitle

\centerline{\textbf{Tweezer array generation}}
\label{sec:setup}
\vspace{1mm} 

\begin{figure}
    \centering
    \includegraphics[width = 9cm]{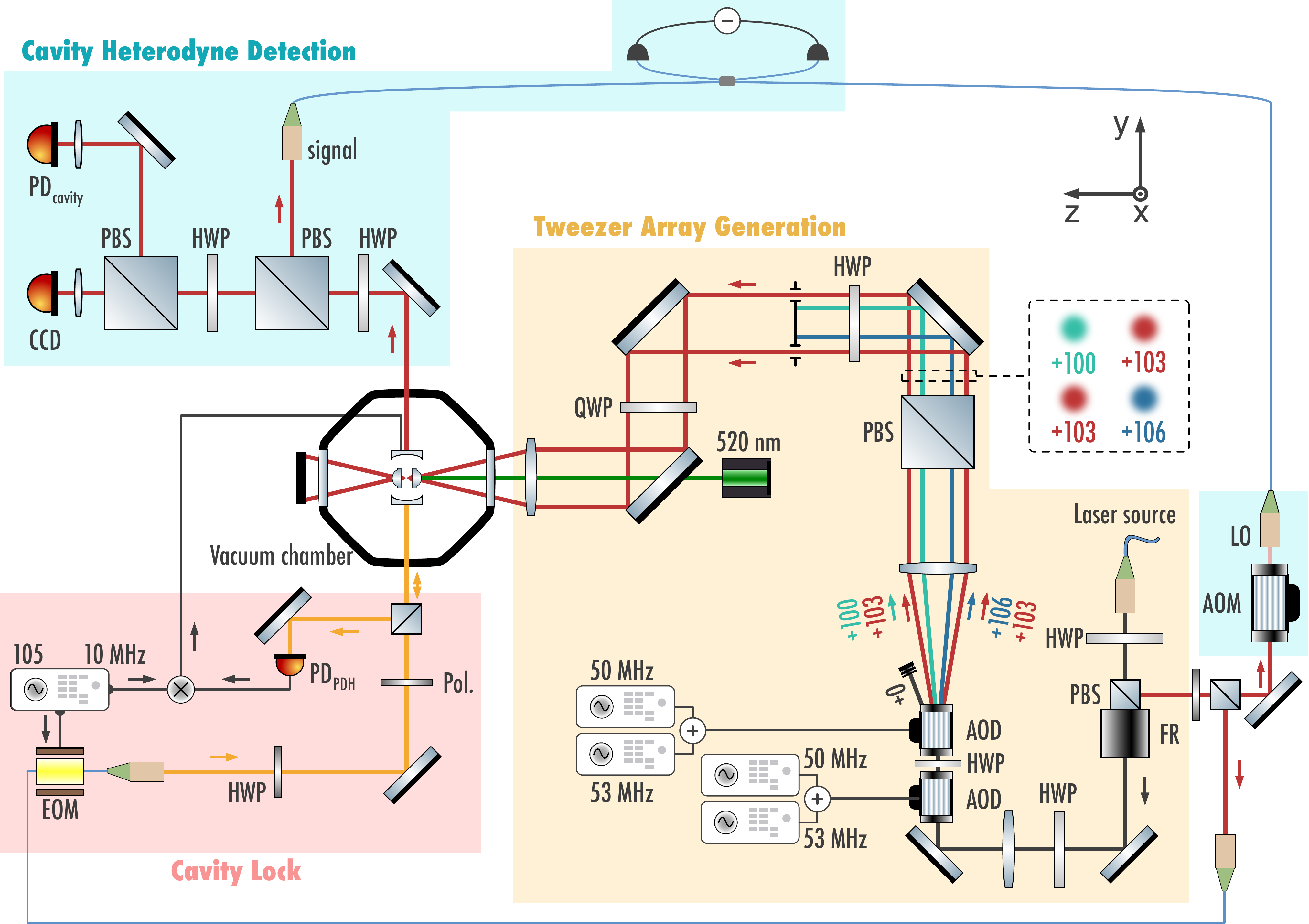}%
    \caption{Figure S1. \textbf{Sketch of full experimental setup.}
    Tweezer arrays are generated using two AODs fed by two RF input channels each (orange box).
    Two beams with the same optical frequency are sent into the vacuum chamber to trap particles.
    A Pound-Drever-Hall scheme is used to lock the cavity (red box), and a balanced heterodyne detection scheme is used to obtain spectral information about mechanical motion (blue box).
    }
    \label{fig:fullsetup}
\end{figure}

A complete sketch of the setup used in this study is shown in Fig.~S\ref{fig:fullsetup}.
The orange box indicates the part of the setup used in the generation of tweezer arrays.
Laser light at $1550\,$nm is sent through two AODs (AA optoelectronics DSTX), that are placed orthogonal to one another.
Each AOD receives the sum of two RF frequencies close to its central frequency, generally one at $50\,$MHz and the other at $53\,$MHz, generated by two different channels of a function generator (MOGLABS Agile RF Synthesizer).
As a result, four optical beams with frequencies $100\,$MHz, $103\,$MHz, $103\,$MHz and $106\,$MHz are generated in the first order following both AODs (see inset of Fig.~S\ref{fig:fullsetup}).
A spatial filter is then used to remove the $100\,$MHz and $106\,$MHz beams, along with any other zero order beams, resulting in only the two diagonal beams at $103\,$MHz making it to the high NA lens inside the chamber.
The two orthogonal AODs are mounted at an angle of $45\deg$ with respect to the optical table, ensuring that the two $103\,$MHz beams are parallel to the cavity.
The RF amplitudes of all four channels are normally kept at $32\,$dBm.
The RF frequencies and amplitudes are fully programmable, a feature that is heavily used in our experiments - the tunability of the frequency allows us to position the particles along the cavity, and the tunability of the amplitude allows us to perform sweeps of their motional frequencies.

\vspace{5mm} 
\centerline{\textbf{Tweezer positioning and calibration}}
\label{sec:position}
\vspace{1mm} 

\begin{figure}
    \centering
    \includegraphics[width = 9cm]{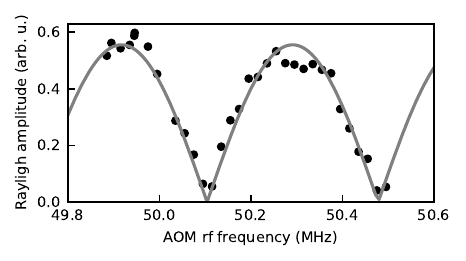}
    \caption{Figure S2. \textbf{Calibration of RF to displacement.}
    The periodicity of measured carrier amplitude extracted by a fit (line) gives the conversion factor between rf and particle displacement.
    }
    \label{fig:positionscan}
\end{figure}



In our investigation of distance-dependence of the interactions, we change the inter-particle separation while keeping the tweezers at the same frequency.
The frequencies of one RF channel of both tweezers are increased (decreased), leading to both tweezers moving further apart (closer together). 
We then move both particles common-mode using a linear translation stage (Attocube ANP(x,y,z)101), in order to bring particle $1$ back to its initial position, in this case a node of the cavity.
To confirm the position at the node we momentarily introduce a frequency difference of $1\,$kHz between the two tweezers to separate their signatures in the spectrum and move the Attocube stage, if required, to minimise the Rayleigh scattered light from particle $1$.
Finally, the cavity detuning is adjusted to the new optical frequency common to both tweezers.
Repeating these steps we change the inter-particle distance while keeping one particle at a node and the detuning fixed.
This procedure also shifts both particles perpendicular to the cavity axis. We can neglect this effect as the displacements of few micrometers are far smaller than the cavity waist $W_\mathrm{c}=50\,\rm{\mu}$m.
Alternatively, if we use both channels of the AODs that go to each tweezer, we can move both particles only along the cavity axis, or only perpendicular to it while changing the tweezers frequency.

To calibrate the spatial separation we sweep the input rf and observe the height of the Rayleigh peak of a single particle on the heterodyne detector.
Fig.~S\ref{fig:positionscan} shows the Rayleigh amplitude $I_\text{RL}\propto |\cos (\varphi)|$ scales with the position $\varphi$ in the standing wave, which is periodic with $\lambda/2$.
By extracting the periodicity of the fit (line) to the data points we get the conversion factor $(1.34\pm0.02)\lambda/\,$MHz of rf to displacement. Since the tweezers' position scales linearly with the rf applied to the AODs, we can use this conversion factor to obtain the usual inter-particle distance for rf inputs of $50\,$MHz and $53\,$MHz of $(4.01\pm0.05)\lambda\approx 6\,\mathrm{\mu}$m. 

\vspace{5mm} 
\centerline{\textbf{Coherent scattering with two nanoparticles}}
\label{sec:theory}
\vspace{1mm} 

The theory for coherent scattering of a single nanoparticle can be found in~\cite{GonzalezBallestero2019}. 
Here we briefly lay out the necessary extensions for two nanoparticles and the emerging effective coupling. We consider two particles each trapped by a different optical tweezers and coupled to an optical cavity. We assume both lasers have the same frequency $\omega_0$ and waist $W_t$, are polarized either parallel or perpendicular to the cavity axis $y$, propagate along $z$, and that their foci are separated by a distance $d\gg W_t$. The dynamics of the system formed by the single cavity mode and the center-of-mass motion of the two particles is governed by the following master equation for its density matrix, $\hat{\rho}$:
\begin{multline}
    \dot{\hat{\rho}} = -\frac{i}{\hbar}\left[\hat{H},\hat{\rho}\right] + \kappa\left(\hat{c}\hat{\rho}\hat{c}^\dagger-\frac{1}{2}\left\{\hat{c}^\dagger\hat{c},\hat{\rho}\right\}\right)
    \\
    -\sum_{i=1,2}\sum_{\mu=x,y,z}\frac{\Gamma_{i\mu}}{2}\left[\hat{b}_{i\mu}+\hat{b}_{i\mu}^\dagger,\left[\hat{b}_{i\mu}+\hat{b}_{i\mu}^\dagger,\hat{\rho}\right]\right]
   \\ +\sum_{i=1,2}\sum_{\mu=x,y,z}\frac{\gamma_{i\mu}}{4}\left[\hat{b}_{i\mu}+\hat{b}_{i\mu}^\dagger,\left\{\hat{b}_{i\mu}^\dagger-\hat{b}_{i\mu},\hat{\rho}\right\}\right]
\end{multline}
where $\kappa$ is the cavity linewidth, $\hat{c}$ is the anihilation operator of a cavity photon, and the curly brackets denote the anticommutator. The motional mode $\mu$ of particle $i$ is characterized by an annihilation operator $\hat{b}_{i \mu}$, a friction rate $\gamma_{i\mu}$, and a heating rate $\Gamma_{i\mu}$ which includes contributions from surrounding gas molecules and from laser shot noise. In a frame rotating at the frequency of the two optical tweezers $\omega_0$ the coherent scattering Hamiltonian is
\begin{equation}\begin{aligned}
\hat{H}&=\hbar \sum_{i=1,2} \sum_{\mu=x, y, z} \Omega_{i, \mu} \hat{b}_{i \mu}^{\dagger} \hat{b}_{i \mu}\\
&+\hbar \Delta \hat{c}^{\dagger} \hat{c}+\hbar\left(\Omega_c \hat{c}^{\dagger}+\Omega_c^* \hat{c}\right)\\
&+\hbar \sum_i \sum_\mu\left[\hat{c}^{\dagger} g_{i \mu}\left(\hat{b}_{i \mu}^{\dagger}+\hat{b}_{i \mu}\right)+\text { H.c. }\right]\,,
\end{aligned}\end{equation}
with $\Delta = \omega_c-\omega_0$, $g_\alpha$ the coherent scattering couplings, and with a cavity driving given by
\begin{equation}
    \Omega_c \equiv -\frac{1}{2}\sqrt{\frac{\omega_c}{2\hbar\varepsilon_0 V_c}}\sum_j\alpha_jE_{0j}\cos\theta_j\cos\varphi_j e^{-i\Phi_j}.
\end{equation}
Here, $\omega_c$ and $V_c$ are the cavity bare frequency and mode volume, $\alpha_j$ is the polarizability of particle $j$, $E_{0j}$ and $\theta_j$ the electric field amplitude and polarization angle of tweezer $j$ at its focus ($\theta_j=0$ for polarization perpendicular to cavity axis, $\theta_j=\pi / 2$ for polarization along cavity axis), and $\Phi_j$ the phase of each trapping laser, which we choose for convention as $\Phi_1=0, \Phi_2=\Phi$. The angle $\varphi_j$ encodes the position of the focus of tweezer $j$ within the cavity mode profile, specifically $\varphi_j=0$ at an antinode and $\varphi_j=\pi/2$ at a node. For $\vert\Omega_c / \kappa\vert \ll 1$ the coherent scattering couplings read
\begin{equation}\label{eq:optomechanical_coupling}
\begin{array}{r}
{\left[\begin{array}{c}
g_{j x} \\
g_{j y} \\
g_{j z}
\end{array}\right]=\sqrt{\frac{\omega_c}{2 \hbar \epsilon_0 V_c}} \frac{\alpha_j E_{0 j}}{2} \cos \left(\theta_j\right) e^{-i \Phi_j}} \\
\times\left[\begin{array}{c}
(-1)^j k_c x_{0 j} \sin \varphi_j \sin \theta_j \\
(-1)^j k_c y_{0 j} \sin \varphi_j \cos \theta_j \\
-i k_0 z_{0 j} \cos \varphi_j
\end{array}\right]
\end{array}\,,
\end{equation}
with $k_c$ and $k_0$ the wavenumber of cavity and tweezer modes and $\left\{x_{0 j}, y_{0 j}, z_{0 j}\right\}$ the zero-point motion along the three axes. This expression also assumes that the cavity Rayleigh range is much smaller than the separation between particles, which is the case for this experiment as $y_\mathrm{R} = \pi W_\mathrm{c}^2/\lambda_0 = (5\pm 1)\,$mm $\ll d$ (with $W_c$ the cavity waist).

To compute the eigenfrequencies of the coupled system, i.e. the peaks of the cavity power spectral density, it is sufficient to compute the dynamics of the classical motional amplitudes, defined by the vector
\begin{equation}
\mathbf{v}\equiv\left[\left\langle\hat{b}_{1 x}\right\rangle,\left\langle\hat{b}_{1 y}\right\rangle,\left\langle\hat{b}_{1 z}\right\rangle,\left\langle\hat{b}_{2 x}\right\rangle,\left\langle\hat{b}_{2 y}\right\rangle,\left\langle\hat{b}_{2 z}\right\rangle\right].
\end{equation}
We compute their dynamics by first adiabatically eliminating the cavity~\cite{WilsonRaeNJP2008,GonzalezBallestero2023tutorial} to obtain a reduced master equation for the motional modes. This procedure is valid when $\kappa \gg \vert g_{i\mu}, \gamma_{i\mu}\vert$ and amounts to considering the cavity a passive bath that couples the motional modes. Then, we use the resulting master equation to compute the dynamics of the vector of mechanical amplitudes, given in the underdamped regime $\vert \gamma_{i\mu}\vert \ll \Omega_{i,\mu}$ by
\begin{equation}
\frac{d}{d t} \mathbf{v}=-i \bar{A} \mathbf{v},
\end{equation}
with a dynamical matrix
\begin{equation}
\bar{A}_{\alpha \alpha^{\prime}} \equiv\left[\Omega_\alpha-i \frac{\gamma_\alpha}{2}\right] \delta_{\alpha \alpha^{\prime}}+G_{\alpha \alpha^{\prime}}\,,
\end{equation}
where we defined the multi-index $\alpha=\{j=(1,2), \mu=(x, y, z)\}$. The effective cavity-mediated couplings read
\begin{equation}\label{eq:cavity_coupling}
G_{\alpha \alpha^{\prime}}=\frac{g_{\alpha^{\prime}}^* g_\alpha}{\left(\Delta+\Omega_{\alpha^{\prime}}\right)+i \kappa / 2}+\frac{g_{\alpha^{\prime}} g_\alpha^*}{\left(\Delta-\Omega_{\alpha^{\prime}}\right)+i \kappa / 2}\,.
\end{equation}
Note that aside from couplings between different modes the cavity also induces a mechanical frequency shift $G_{\alpha\alpha}$ in mode $\alpha$. Note also that the couplings are in general non-reciprocal, i.e., $G_{\alpha\alpha'} \ne G_{\alpha'\alpha}$.

The peaks of the PSD will be centered at frequencies $\operatorname{Re}\left[\Lambda_l\right]$ and have linewidth $\operatorname{Im}\left[\Lambda_l\right]$, where $\Lambda_l$ are the six eigenvalues of the matrix $\bar{A}$. All mechanical modes that do not couple to the cavity ($g_\alpha=0$) remain uncoupled in this effective picture and give rise to trivial eigenvalues $\Lambda_l = \Omega_\alpha - i\gamma_\alpha/2$. In the experiment we consider purely $x-$polarized tweezers so that both $x-$mechanical modes are uncoupled from the cavity. If we particularize to the case where both particles are at the node ($\varphi_1=\varphi_2=\pi/2$) or at the antinode ($\varphi_1=\varphi_2=\pi$) two more mechanical modes are uncoupled from the cavity, namely the $y-$modes at the node and the $z-$modes at the antinode. In these cases the remaining two mechanical modes $\alpha=\{1,\mu\}$ and $\alpha'=\{2,\mu\}$ form a $2\times2$ coupled system which can be diagonalized analytically to obtain the following two eigenvalues,
\begin{equation}
\begin{aligned}
\Lambda^{ \pm}_\mu=\frac{1}{2} & {\left[D_1(\Delta)+D_2(\Delta)\right.} \\
& \left. \pm \sqrt{\left(D_1(\Delta)-D_2(\Delta)\right)^2+4 G_{1 \mu 2 \mu} G_{2 \mu 1 \mu}}\right]
\end{aligned}
\end{equation}
with
\begin{equation}
D_j(\Delta) \equiv \Omega_{j, \mu}-i \frac{\gamma_{j \mu}}{2}+G_{j \mu j \mu} .
\end{equation}
The normal mode frequencies $\lambda_\mu^\pm$ correspond to the real part of $\Lambda_\mu^\pm$, and their difference corresponds to the mode anticrossing
\begin{equation}
\label{eq:splitting}
\lambda^{+}_\mu-\lambda^{-}_\mu=\operatorname{Re} \sqrt{\left(D_1(\Delta)-D_2(\Delta)\right)^2+4 G_{1 \mu 2 \mu} G_{2 \mu 1 \mu}}.
\end{equation}
At the point of avoided crossing, where $\Omega_{1,\mu}=\Omega_{2,\mu}$, for identical particles, and for in-phase tweezers ($\Phi_2=0$), the cavity-mediated couplings become reciprocal and the anticrossing simplifies to $2\sqrt{G_{1\mu,2\mu} G_{2\mu,1\mu}} \approx 2\abs{G_{1\mu,2\mu}} \equiv 2\abs{G_{\mu,\mu} }$ as used in the main text.

\vspace{5mm} 

\centerline{\textbf{Extracting mode splitting}}
\label{sec:fitting}
\vspace{1mm} 

To extract the mode splittings $\min(|\lambda^{+}_\mu-\lambda^{-}_\mu|)$ from the shapes in our measured spectrograms we use Eq.~\ref{eq:splitting}. 
The normal mode frequencies are dependent on cavity parameters $\kappa$ and $\Delta$, which we independently measure, the power $P_i$ in each of the tweezers, and the bare mechanical frequencies $\Omega_{i,\mu}$.  
First, we run a peak finding algorithm for each slice of the spectrograms and sort the peaks into their respective modes.
We then extract $\Omega_{i,\mu}$ from the edges of the spectrograms.
The uncoupled $x$ peaks give a calibration for the relative powers by fitting a square-root function ($P_i \propto \sqrt{\Omega_{i,x}}$) to them. 
Finally, we subtract the peaks corresponding to the modes we want to fit from each other and fit Eq.~\ref{eq:splitting}, inserting the cavity parameters, powers and bare frequencies. The only fit parameter left is the product of optomechanical couplings $g_{1,\mu} g_{2,\mu}$. 
The mode splittings are then presented as the minimal separation between the fitted lines $\operatorname{min}(|\lambda^{+}_\mu-\lambda^{-}_\mu|)$.
For strongly coupled modes (for example in Fig. 4d where $G_{zz}/\Omega_{z} = 0.24$) the fidelity of the fit is reduced. The strong coupling case and the cooling of such strongly hybridised modes will be the topic of a future study.

\vspace{5mm} 
\centerline{\textbf{Estimations of coupling strengths}}
\label{sec:coupling}

\begin{table}
    \centering
    \begin{tabular}{r|c !{\vrule width 2pt} r|c}
$P_\mathrm{t}$ & $(0.13\pm0.01)\,$W                   &$\omega_\mathrm{c}$ & $\omega_0 + \Delta$  \\
$W_\mathrm{t}$ & $(0.85\pm 0.1) \times 10^{-6}\,$m  &$W_\mathrm{c}$ & $(50\pm 5)\times10^{-6}\,$m\\
$\epsilon_0$ & $8854\times10^{-15}\,$F/m            &$L_\mathrm{c}$ & $(9.6\pm 0.1)\times10^{-3}\,$m\\
$\lambda_0$ & $1550\times10^{-9}$\,m                &$\theta$ & $(0\pm5)\pi/180$\\
$R$ & ($75\pm 4)\times10^{-9}\,$m                     &$\epsilon$ & $2.07$\\
$\rho$ & $2200\,$kg/m$^3$                           &$\alpha$ & $4\pi\epsilon_0 R^3 (\epsilon-1)/(\epsilon+2)$\\
    \end{tabular}
    \caption{Table 1. Experimental parameters and their uncertainties used for estimating coupling rates. $P_\mathrm{t}$: optical powers of both tweezers at the avoided crossing, $W_\mathrm{t}$: waist of the tweezers, $R$: nominal radius of the nanoparticles, $\rho$: density of the nanoparticles, $\omega_\mathrm{c}$: frequency of the cavity field, $W_\mathrm{c}$: cavity waist, $L_\mathrm{c}$: cavity length, $\epsilon$: electrical susceptibility of the nanoparticles, $\alpha$: polarisability of the nanoparticles}
    \label{tab:parameters}
\end{table}
For all estimations of coupling strengths we use the system parameters in Table~\ref{tab:parameters}. $\Delta$ and $\kappa$ are taken from the main text and we set $\varphi$ via from the position in the cavity.
From our parameters and with Eq.~\ref{eq:optomechanical_coupling} we estimate maximal optomechanical coupling strengths of $g_{i,y}/2\pi = (32\pm4)\,$kHz and $g_{i,z}/2\pi = (50\pm7)\,$kHz for $\varphi = \pi/2$ and $\varphi = 0$, respectively. 
We then get the effective cavity-mediated couplings calculated from Eq.~\ref{eq:cavity_coupling} and present them including uncertainties in the system parameters as shaded areas in all plots.

Using the same system parameters we estimate the coupling strengths of direct optical and Coulomb interactions from
\begin{align}
|G_\text{OB}|&=\frac{\alpha^2 k_0^5 P_\mathrm{t}}{4 c W_\mathrm{t}^2 \pi^2 \epsilon_0^2 m \Omega} \frac{\cos \left(k_0 d\right)}{k 0 d}\,,\\
|G_\text{C}| &= \frac{Q_1 Q_2}{8 \pi \epsilon_0 d^3} \frac{1}{m \sqrt{\Omega_1 \Omega_2}}\,.
\end{align}
We make the conservative assumption of $Q_1 = Q_2 = 50 e$ as nanoparticles of our size, loaded by spraying a fine solution into the trapping volume, typically hold few tens of elementary charges. 
For the minimum inter-particle distance of {\color{black}$d/\lambda = 3.5$ in Fig.~3 we obtain $G_\text{C}/2\pi = (0.17\pm0.03)\,$kHz and $G_\text{O}/2\pi = (0.14\pm0.04)\,$kHz, respectively.} 
Typical peak widths broadened by cavity cooling in our experiments are $(0.6\pm0.1)\,$kHz. 
Assuming we can resolve peaks separated by half their widths, we estimate a minimum measurable coupling strength of $G_\mathrm{min}/2\pi \sim 0.15\,$kHz close to both $G_\text{C}$ and $G_\text{O}$.

\vspace{5mm} 
\bibliography{Supplementary}